\documentclass{article}

\usepackage{microtype}
\usepackage{graphicx}
\usepackage{subfigure}
\usepackage{booktabs} 

\usepackage{hyperref}

\usepackage[accepted]{icml2021}

\icmltitlerunning{Graph Representation Learning on Tissue-Specific Multi-Omics}

\begin{document}

\twocolumn[
\icmltitle{Graph Representation Learning on Tissue-Specific Multi-Omics}



\icmlsetsymbol{equal}{*}

\begin{icmlauthorlist}
\icmlauthor{Amine Amor}{to}
\icmlauthor{Pietro Lio'}{to}
\icmlauthor{Vikash Singh}{to}
\icmlauthor{Ramon Viñas Torné}{to}
\icmlauthor{Helena Andres Terre}{to}
\end{icmlauthorlist}

\icmlaffiliation{to}{Department of Computer Science and Technology, University of Cambridge, Cambridge, United Kingdom.}

\icmlcorrespondingauthor{Amine Amor}{amine.amor98@gmail.com}


\icmlkeywords{Machine Learning, Graph Representation Learning, Variational Graph Auto-Encoders, Generative Models, Link Prediction, Multi-Omics, ICML}

\vskip 0.3in
]



\printAffiliationsAndNotice{}  

\begin{abstract}
Combining different modalities of data from human tissues has been critical in advancing biomedical research and personalised medical care. In this study, we leverage a graph embedding model (i.e VGAE) to perform link prediction on tissue-specific Gene-Gene Interaction (GGI) networks. Through ablation experiments, we prove that the combination of multiple biological modalities (i.e multi-omics) leads to powerful embeddings and better link prediction performances.
Our evaluation shows that the integration of gene methylation profiles and RNA-sequencing data significantly improves the link prediction performance.
Overall, the combination of RNA-sequencing and gene methylation data leads to a link prediction accuracy of 71\% on GGI networks. By harnessing graph representation learning on multi-omics data, our work brings novel insights to the current literature on multi-omics integration in bioinformatics. 
\end{abstract}

\section{Introduction}

Thorough understanding of human health and pathological conditions requires the analysis of molecular data at different levels, such as genome, epigenome, transcriptome, proteome, and metabolome. 
To account for the interactions between these omics and study complex biological processes holistically, it is fundamental to follow an integrative approach which combines multi-omics (i.e multiple modalities of biological data) \cite{multi-omics-review}. Integrative approaches help to evaluate the flow of information from one omic layer to another, and therefore contribute to bridge the gap between genotype and phenotype.
In the era of precision medicine, high-throughput technologies can generate very large amounts of multi-omics data, and contribute to improve prognostics of disease phenotypes. While there has been a significant interest in building integrative systems in bioinformatics \cite{Computational-Oncology}, multi-omics integration on tissue-specific data has been underexplored. Motivated by the lack of research on tissues functional diversity, we focus our study on tissue-specific biological data using 3 modalities (i.e omics): Gene-Gene Interaction networks (GGI), RNA sequencing data and gene methylation profiles. Therefore, our input consists of a network of interacting genes with their gene expression features (i.e RNA sequencing and gene methylation).

Overall, the novelty of our work relies on the analysis of tissue-specific data and the integration of multi-omics features using a graph embedding model (VGAE). 

\section{Related Work}

\subsection{Tissue-specific research}

The heterogeneity of cells across tissues is a major challenge for understanding biological processes and developing therapeutic targets of distinct tissues. Although tissue-specific mechanisms are rarely explored, there have been research initiatives to identify tissue-specific molecular profiles. Jambusaria et al. \cite{cellular-heterogeneity-tissues} developed a predictive model called ``HeteroPath" which produces unique tissue-specific gene regulatory networks. By identifying distinct cellular populations in tissue transcriptomic datasets, ``HeteroPath" contributes to improve the comprehension of tissue-specific phenotypes.
Whereas this study focuses on transcriptomics, metabolomics have also been investigated in the context of tissue-specific analysis. For instance, CORDA \cite{CORDA} (Cost Optimization Reaction Dependency Assessment) is a genome scale model that detects important metabolic reactions across various human tissues. Using CORDA algorithm, the authors developed 76 healthy and 20 cancer tissue-specific reconstructions, and identified metabolic pathways shared across tissues.

We notice that these papers explore metabolomics and transcriptomics independently, to infer molecular signatures of tissues. Motivated by the potential complementarity of omics features, our approach incorporates diverse modalities of omics to provide a more global molecular perspective of distinct human tissues.


\begin{figure*}[h]
  \includegraphics[width=\textwidth]{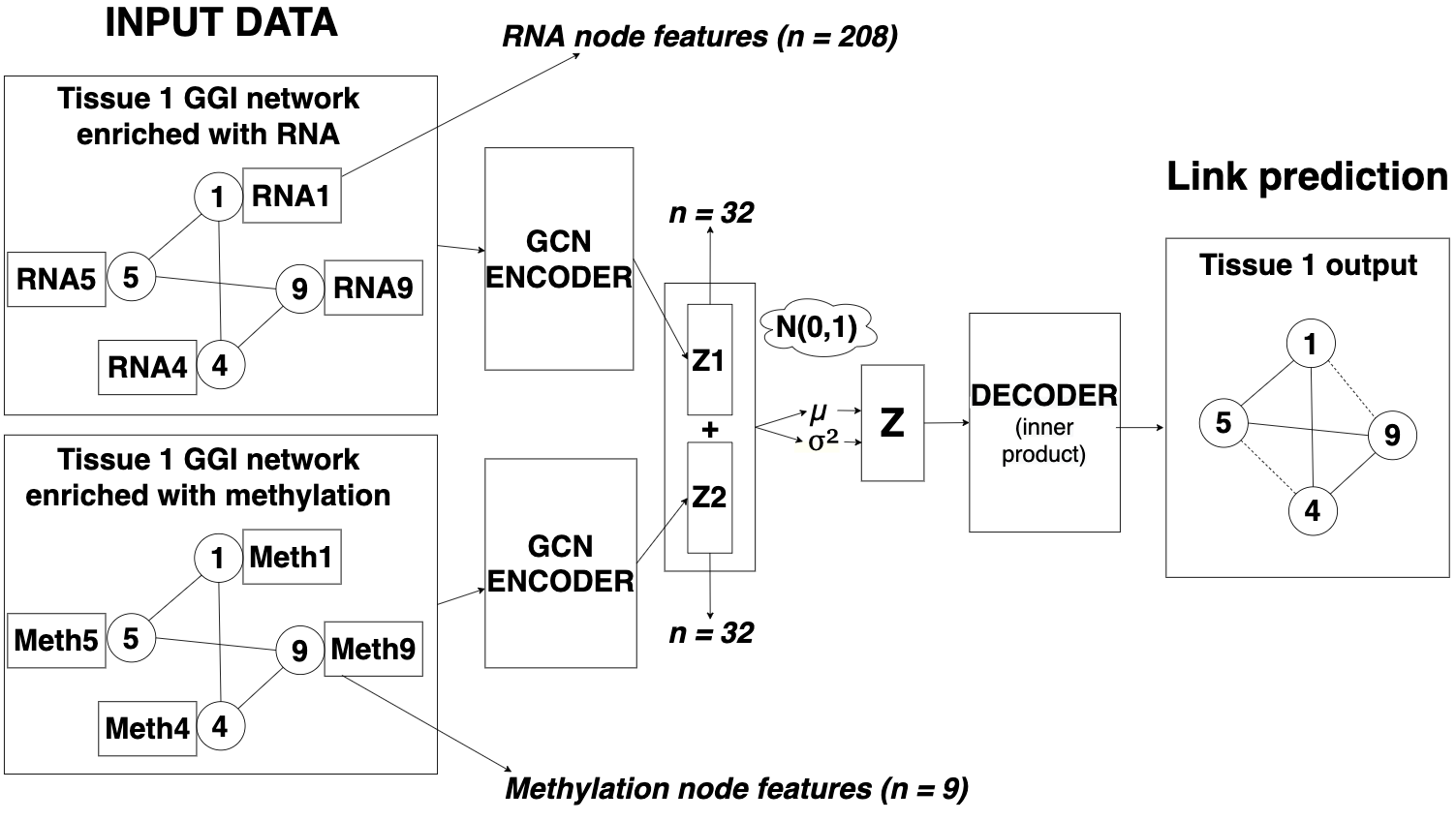}
  \caption{A \textbf{VGAE} (adapted from \cite{VGAE_original_paper}) that performs an intermediate integration of gene methylation and RNA sequencing features. Both input networks are tissue-specific Gene-Gene Interaction (GGI) networks. While they have the same adjacency matrix (from Tissue 1 GGI network), they are enriched with different node features (i.e gene methylation or RNA sequencing features).}
  \label{fig:GVAE-fig}
\end{figure*}

\subsection{Graph representation learning on tissue-specific expression data}

Biological processes can be described in terms of molecular interactions that occur across multiple omics layers. This type of data comes in the form of interaction networks, which have been used to train several graph embedding models on the prediction of gene-disease associations \cite{gene-disease-embeddings} \cite{gene-disease-pietro-Vikash} and the identification of molecular signatures \cite{GDL-PPInetworks}.

Regarding tissue-specific analysis, Ohmnet \cite{ohmnet} is an unsupervised node feature learning framework which predicts multicellular function through multi-layer tissue protein-protein interaction (PPI) networks. It represents one of the rare initiatives that uses graph embedding techniques on tissue-specific molecular interactions. 

Overall, substantial research was conducted on multi-omics integration frameworks and graph representation learning. However, to the best of our knowledge, multi-omics integration on tissue-specific graphs/networks is a research area that is relatively poor. Therefore this study leverages graph representation learning on tissue-specific multi-omics data.

\section{Multi-omics integration with VGAE}
\label{inter-VGAE-model}

\subsection{Data collection}

We collect tissue-specific GGI networks, RNA sequencing data and gene methylation profiles from 3 public databases:

\begin{itemize}
    \item \textbf{HumanBase (GIANT):} It is a public database that provides human genomic data such as gene expression, regulation and interaction networks. From HumanBase, we collect 5 tissue-specific Gene-Gene Interaction (GGI) networks \cite{multicellular_function_tissue}, which were built using gene expression and gene function from a large compendium of tissues and cell-types.
    
    \item \textbf{The Genotype-Tissue Expression (GTEx) project:} Launched by the National Institutes of Health (NIH) in September 2010, the Genotype-Tissue Expression project (GTEx \cite{GTex}) is a public resource that gives access to tissue-specific gene expression and regulation data. The samples were collected from 54 healthy tissue sites across nearly 1000 participants. From GTEx, we download 5 tissue-specific filtered and normalised gene expression matrices (RNA sequencing data). 
    
    \item \textbf{MethBank 3.0:} MethBank \cite{Methbank} is a public database that was developed in 2017 by the Big Data Center of Beijing Institute of Genomics. The database incorporates 34 consensus reference methylomes derived from 4,577 healthy human samples at different ages. From MethBank, we collect normalised healthy human gene methylation profiles for 5 tissues.
\end{itemize}

\subsection{Variational Graph Auto-Encoder (VGAE)}

To perform link prediction on tissue-specific GGI networks, we employ an unsupervised variational graph autoencoder (VGAE) \cite{VGAE_original_paper} that integrates distinct latent representations derived from RNA sequencing data and gene methylation profiles (i.e Z1 and Z2 in Figure \ref{fig:GVAE-fig}). The combined representation is fed into the decoder of the VGAE which aims to reconstruct the adjacency matrix of the original network. The reconstruction of an adjacency matrix is also known as the link prediction problem. In the reconstruction output (in Figure \ref{fig:GVAE-fig}), solid lines (positive edges) represent the existence of a link between 2 nodes, whereas dotted lines (negative edges) represent the absence of link.
In our study, we train our integrative VGAE on tissue-specific Gene-Gene Interaction networks (GGI) where nodes represent genes and edges/links represent a functional interaction between genes.
In Figure \ref{fig:GVAE-fig}, the boxes represent the feature vectors associated with the genes in the adjacency matrix. ``Meth'' represent gene methylation features whereas ``RNA'' represent RNA-sequencing features.

As shown on Figure \ref{fig:GVAE-fig}, there is a significant gap of dimensionality between RNA sequencing features (n=208) and gene methylation features (n=9). Indeed, for each gene, there is a vector of 208 RNA sequencing features and a smaller vector of 9 methylation features. In order to preserve the unique distribution of each data type, the integrative VGAE combines the features representations in the latent space, rather than the input space. The first step consists of training two separate GCN (Graph Convolutional Neural Network) encoders on a GGI network enriched with RNA sequencing data and a GGI network enriched with gene methylation data, respectively. The GCNs encode the features into 2 separate embeddings Z1 and Z1, which have the same dimensions (n=32).
Z1 and Z2 are then concatenated and fed into the rest of the VGAE which performs link prediction. Since Z1 and Z2 have the same shape, this approach gives the same weight to methylation and RNA sequencing features, despite their initial imbalance of dimensions. 
Additionally, unlike an early integration approach which combines features at the input level, our model does not increase the dimensionality of the input space. However, our intermediate integration requires to train an additional GCN encoder and therefore increases the number of parameters to learn.

\section{Evaluation and Results}

The experiments aim to evaluate how much each omics data contributes to the performance of the models. To that end, we conduct an ablation study which consists of combining multi-omics in three different ways: GGI+RNA, GGI+Meth, GGI+RNA+Meth. The ablation study helps to assess the individual importance of each data modality (RNA-sequencing or gene methylation) as well as their complementarity in achieving link prediction. This provides biological insights into the relevance of particular omics in learning tissue-specific representations.

Additionally, we compare the performance of the VGAE to the non-generative Graph Auto-Encoder (GAE) in order to understand the relevance of generative models in multi-modal learning on graphs.

\subsection{Multi-omics integration results}

The table illustrates the average link prediction performance of the VGAE on 5 tissue-specific GGI networks. Here, ``Bal Acc'' refers to the balanced accuracy metric defined as $(TPR + TNR)/2$.

\begin{center}
\begin{tabular}{lll} \toprule
    {\textbf{Integration}} & {\textbf{Bal Acc}} & {\textbf{F1 score}}  \\ \midrule
    GGI & $51 (\pm 1.9)$ & $33 (\pm 2.5)$ \\
    GGI+Meth &  $56.7 (\pm 3.8)$ & $48.0 (\pm 4.5)$ \\
    GGI+RNA &  $71.4 (\pm 3.9)$ & $69.6 (\pm 2.2)$ \\
    GGI+RNA+Meth & $\textbf{71.6} (\pm 6.1)$ & $\textbf{70.1} (\pm 6.7)$ \\ \bottomrule
\end{tabular}
\end{center}

\subsection{Generative vs Non-Generative models results}

The table shows the average link prediction performance of the VGAE and GAE on 5 tissue-specific GGI networks, using the intermediate integration approach described in section \ref{inter-VGAE-model}.

\begin{center}
\begin{tabular}{lll} \toprule
    {\textbf{Model}} & {\textbf{Bal Acc}} & {\textbf{F1 score}}  \\ \midrule
    GAE  & $70.2 (\pm 3.0)$ & $68.1 (\pm 4.2)$ \\
    VGAE & $\textbf{71.6} (\pm 4.1)$ & $\textbf{70.1} (\pm 4.7)$ \\ \bottomrule
\end{tabular}
\end{center}

\section{Discussion}

\subsection{Multi-omics integration}

We discuss the results obtained from different types of multi-omics integration in order to understand the value of each omics in achieving link prediction on tissue-specific networks.
On non-enriched GGI (Gene-Gene-Interaction) networks, the VGAE achieves a very poor performance, which highlights the importance of node features to learn informative graph embeddings. 
By adding gene methylation node features (GGI+Meth), we observe a notable improvement of the overall performance. The balanced accuracy grows from 50\% to 57\% and the F1 score grows from 33\% to 48\%.  
On the other hand, augmenting the networks with RNA node features (GGI+RNA) brings a considerable enhancement in the link prediction performance. The incorporation of RNA features causes the balanced accuracy and F1 score to increase from 50\% to 70-71\%.
These results suggest that RNA sequencing features are more valuable than gene methylation features and lead to more accurate graph embeddings on tissue-specific GGI networks.
While both RNA and methylation features enhance the prediction performance of the VGAE, their combination (GGI+RNA+Meth) is not particularly complementary for link prediction. Indeed, the VGAE's performance on GGI+RNA+Meth is almost equal to its performance on GGI+RNA.

\subsection{Generative vs Non-Generative models}

On the other hand, we observe that the VGAE results in a higher link prediction performance than the GAE. Indeed, the balanced accuracy and F1 score are roughly 1-2\% higher in the case of the VGAE. 
The higher performance of the VGAE shows the benefits of latent space regularisation. By enforcing the latent distribution to be close to a gaussian distribution, the VGAE regularises the latent space and enables a better generalisation performance. Moreover, the VGAE provides flexibility in the learning process because we can tune the KLD loss with a parameter $\alpha$ and the reconstruction loss with a parameter $\beta$. Increasing $\alpha$ would augment the generative power of the VGAE whereas increasing $\beta$ would further optimise the reconstruction performance.
Overall, these results highlight the relevance of generative models in performing multi-modal learning on multi-omics networks.

\section{Conclusion}

In summary, our work explores multi-modal learning on tissue-specific gene-gene interaction (GGI) networks. Our approach towards multi-omics integration consists of enriching GGI networks with RNA-sequencing and gene methylation features. Since omics modalities are collected separately across distinct tissues, our data is tissue-specific. In order to learn powerful molecular representations, we decide to leverage graph embedding models (i.e VGAE) which have the benefit of being scalable to the incorporation of multiple omics modalities. 
By evaluating our VGAE model on the addition and the removal of omics features, we conduct an ablation study that provides insights into the benefits of each omics data type (i.e RNA-sequencing and gene methylation).

We observe that the performance of the model increases significantly with the integration of gene methylation profiles and RNA features. Additionally, we discover that RNA features lead to a higher improvement than methylation profiles, which suggests that RNA-sequencing data is more insightful for learning tissue-specific molecular signatures.

On the other hand, the VGAE outperforms the non-generative GAE, which reveals the potential of generative models in multi-modal learning on graphs.
Overall, our integrative VGAE achieves a link prediction accuracy of 71\% on the multi-omics networks (GGI+RNA+Meth), which proves its ability to compress high-dimensional biological networks into informative low-dimensional embeddings.

Overall, this study highlights the benefits of multi-omics integration for link prediction on biological networks. Our insights are based on a variational graph auto-encoder (VGAE) which extracts low-dimensional representations from healthy tissue-specific GGI networks. These representations can serve to enrich existing biological datasets and contribute to downstream supervised tasks such as the detection of bio-markers and the identification of tissue-specific diseases.

\section{Future Work}

This study shows novel insights into the benefits of multi-omics integration in bioinformatcs.
For future work, our approach could be leveraged to target a concrete application in prognostics, such as the detection of breast cancer. 

Based on our approach, we could collect multi-omics data from breast tissues and train our VGAE model to distinguish between healthy breast representations and cancer breast representations.
Since our models are scalable and flexible to the integration of heterogeneous omics features, the prediction of breast cancer would only require to change the multi-omics input data. The omics data could be specific to 2 classes: ``Healthy breast tissues'' and ``Diseased breast tissues''. Additionally, the tissue-specific representations learnt on breast cancer could be used by downstream machine learning classifiers to perform more specialised predictions, such as identifying breast cancer molecular subtypes \cite{DIABLO}.

More generally, our proposed VGAE is interdisciplinary and can be harnessed to perform multi-modal learning on any task involving graph structures (e.g social networks and graph recommendation systems).

\nocite{langley00}

\bibliography{references}
\bibliographystyle{icml2021}





\end{document}